# Optimal Strategies for Virus Propagation

Soumya Banerjee[1,2, 3, 4]

[1] Department of Computer Science, University of New Mexico, USA
[2] Ronin Institute, Montclair, USA
[3] Complex Biological Systems Alliance, USA
[4] Broad Institute of MIT and Harvard, USA
soumya.banerjee@ronininstitute.org

**Abstract.** This paper explores a number of questions regarding optimal strategies evolved by viruses upon entry into a vertebrate host. The infected cell life cycle consists of a non-productively infected stage in which it is producing virions but not releasing them and of a productively infected stage in which it is just releasing virions. The study explores why the infected cell cycle should be so delineated, something which is akin to a classic "bang-bang control" or all-or-none principle. The times spent in each of these stages represent a viral strategy to optimize peak viral load. Increasing the time spent in the non-productively infected phase ($\tau_1$) would lead to a concomitant increase in peak viremia. However increasing this time would also invite a more vigorous response from Cytotoxic T-Lymphocytes (CTLs). Simultaneously, if there is a vigorous antibody response, then we might expect $\tau_1$ to be high, in order that the virus builds up its population and conversely if there is a weak antibody response, $\tau_1$ might be small. These tradeoffs are explored using a mathematical model of virus propagation using Ordinary Differential Equations (ODEs). The study raises questions about whether common viruses have actually settled into an optimum, the role for reliability and whether experimental infections of hosts with non-endemic strains could help elicit answers about viral progression.

## 1  Introduction

A normal cell upon infection goes through a life cycle characterized by 2 phases: a stage in which it is producing virions but not releasing it (non-productively infected stage) and a stage in which it is releasing virions into the outside environment (productively infected stage). Hence there is a delay between infection and release of virions. In whatever follows, we denote the time spent in the non-productively infected stage as $\tau_1$ and the time spent in the productively infected stage as $\tau_2$. The time in $\tau_1$ is spent in viral penetration, uncoating of viral core, transcription and assembly.

The number of virions produced over the entire infected cell life cycle is directly proportional to $\tau_1 + \tau_2$. It is asked whether the virus might be trying to maximize this quantity in order to optimize "virulence" (a quantity which shall be concretized shortly). The question of why there need be 2 distinct phases and not just one where virion production and release occur simultaneously, also cries out for explanation. Such forms of delineation are called "bang-bang control" or the all-or-none principle and are characterized by a phase of proliferation and then terminal differentiation, and are frequently encountered in optimal biological systems [1].

If the total length of the infected cell lifetime is a measure of "virulence", we can then set a theoretical upper bound on it and then compare it with its actual value from field measurements. This would give us a qualitative understanding of "how far" the virus can still go in optimizing itself e.g. it can be used to determine if the avian-influenza virus is already as virulent as it can be or is it still sub-optimal.

The rest of the paper is organized as follows: Section 2 discusses arguments for optimization in biological systems and Section 3 introduces the principle of "bang-bang control". The hypotheses and questions are posed in Section 4 and Section 5 outlines the mathematical model. Section 6 contains the results and discussions and concluding remarks are presented in Section 7.

## 2  Optimization in a Biological System

Before commencing with the mathematical analysis we state what our modeling philosophy will be and give some justification for employing such an approach. First of all, there is certainly no *a priori* reason why virus propagation -- or any other biological system-- should operate in an "optimal" fashion. Indeed there is a substantive issue as to whether the notion of "optimality" can be given an operational meaning for many biological systems. Typically, an organism or a virus is forced to cope with a number of competing influences so that an improvement in one direction involves a sacrifice in another. Thus optimality must be interpreted in a broader sense as a "best compromise" solution. Beyond this consideration, however, there are at least two major reasons why a particular biological system might not be performing its function in the most expeditious fashion. First, despite the fact that one tends to think of natural selection as an inherently optimizing process, improvements on existing mechanisms generally proceed by small modifications of existing structures. Thus there is ample opportunity for the system to become trapped in "local" maxima; there may be "nearby" structures with higher fitness but to reach them may require a temporary, but fatal, decrease in overall fitness. Second, while the system may be constantly improving, evolution is a slow and erratic process so that any system we examine may not have had time to optimize under existing selective pressures. Both of these objections may be partially circumvented by restricting attention to systems which appear to have been evolutionarily static for a long time. The mammalian immune system and viruses surely fulfill this criterion.

A virus typically only wants to proliferate in a host only so much as to ensure transmission to another host (an exception is the Ebola virus which kills its host so fast as to prevent propagation to another host). Hence it is trying to optimize the basic reproductive ratio $R_0$ in epidemic models. In vector-borne pathogens, the peak viremia in blood serum is a very good determinant of $R_0$ [2]. Hence, we assume that the virus is trying to optimize the peak viral load in blood serum ($P_v$).

## 3  Bang-Bang Control

In a seminal paper [1], Perelson et al. examined the mammalian immune system and looked at optimal strategies for B-cell proliferation and differentiation. They used control theoretic principles to analyze the minimum time taken by the immune system to eliminate a fixed amount of antigen in the shortest span of time. The problem briefly stated is as follows: given an initial population of B-cells (which secrete antibodies at a modest rate, proliferate into B-cells or differentiate into plasma cells) and plasma cells (which secrete antibodies at a very high rate but do not proliferate), how do you apportion the total population between B-cells and plasma cells? Does the optimal strategy involve proliferation of B-cells followed by differentiation into plasma cells? Or does it involve simultaneous B-cell proliferation and differentiation? The authors showed using optimal control theory that the optimal strategy for B-cells is to go through a stage of proliferation (to build up their population) and then differentiate into plasma cells. Such a control is called "bang-bang" or all-or-none. It is not immediately evident or intuitive that a strategy of simultaneous B-cell proliferation and differentiation is not optimal.

A parallel is drawn between that work and the problem at hand here, where the infected cell also goes through a phase of production of virions followed by a phase of virion release. The reasons for "bang-bang control" in the infected cell system and its implications are explored in the following sections.

## 4  Hypotheses and Questions

This section explores some of the hypotheses proposed and frames some questions.

There are 2 hypotheses about the non-productively infected stage of the infected cell:

<u>Hypothesis 1</u>:  The virus is not trying to optimize the duration of the non-productive infected stage ($\tau_1$). Hence this time is exactly equal to the time required for viral penetration, uncoating of viral core, transcription and assembly. The interpretation is that as soon as the first complete virions is assembled, the infected cell immediately proceeds to release the virion i.e. it switches to the next phase of productive infection. The obvious disadvantage of this strategy is that the amount of virions produced would be reduced, compared to an approach in which $\tau_1$ is increased. Clearly this strategy is sub-optimal and we do not explore it further.

Hypothesis 2: The virus is trying to optimize peak viral load and hence viral production. However it cannot increase the duration of the productive infected stage ($\tau_2$). This is so because there are physiological limits imposed by the area and strength of the cell wall, which will constrain the duration of virion release. After a threshold, the cell wall will simply fall apart. It can only increase the duration of the non-productive infected stage ($\tau_1$).

(a) Having a high $\tau_1$ would imply an increased virion release count. However, this would come at the cost of increased susceptibility to lysis by Cytotoxic T-Lymphocytes (CTLs). A lower $\tau_1$ would reduce the susceptibility to CTL mediated lysis at the expense of a reduced virion count.

(b) The virus might optimize itself such that it bursts early in the face of a weak antibody response. Conversely, it could burst later (after building up a pool of virions) when confronted with a vigorous antibody response.

Question 1: Why cannot $\tau_1$ increase indefinitely?
Question 2: Why is the optimal control "bang-bang"?

## 4 Mathematical Formulation

A standard mathematical model of virus propagation adapted from Baccam et al. [3] is constructed to test the hypothesis. Ordinary Differential Equations (ODEs) are used to represent populations of virus, infected cells and normal cells. The equations are shown below:

$$\begin{cases} \dfrac{dT}{dt} = -\beta TV \\ \dfrac{dI_1}{dt} = \beta TV - kI_1 - \omega_{CTL} I_1 \\ \dfrac{dI_2}{dt} = kI_1 - \delta I_2 \\ \dfrac{dV}{dt} = pI_2 - cV \end{cases}$$

where  $T$ = target cell population,
    $I_1$ = non-productively infected cell population
    $I_2$ = productively infected cell population

*V* = virus population
*β* = rate constant of infection
*k* = rate of death of non-productively infected cells
*δ* = rate of death of productively infected cells
$\omega_{CTL}$ = rate of CTL-mediated lysis of non-productively infected cells
*p* = number of virions produced per productively infected cell per time step
*c* = rate of clearance of free virus particles

We also get $\tau_1 = 1/k$ and $\tau_2 = 1/\delta$

In this simple ODE model, the population of target cells (normal and uninfected cells) are represented by the variable ***T***. They are also lost due to infection, which is represented by the term *-βTV*. The non-productively infected cells (***I₁***) are supplied by the loss from the target-cell pool and die at a rate proportional to their number density with constant of proportionality ***k***. They are also lysed by (Cytotoxic T-Lymphocytes) CTLs at a rate proportional to their density and with a constant of proportionality of $\omega_{CTL}$. Productively infected cells (***I₂***) are replenished from the non-productive pool and die at a rate proportional to their density and with a constant of proportionality of ***δ***. New virions (***V***) are produced by infected cells at the rate *pI₂* and virions are lost at a rate proportional to the virus concentration with constant of proportionality *c* (representing antibody-mediated virion clearance).

The variation in $\omega_{CTL}$ has been modelled in a time-dependent fashion. Namely it is made to mimic the clonal expansion of a pool of effector CTLs after day 4.

$$\omega_{CTL}(t) = \begin{cases} 0, t < 4 \\ \Omega \times e^{\Theta \times (t-4)}, t \geq 4 \end{cases}$$

The model was parameterized from a study of experimental infection of Influenza A virus in humans [3]. The model was implemented in the Berkeley Madonna package [14] and the code is freely available for download [15]. The model parameters are shown below

| Parameters | $\beta$ [(TCID$_{50}$/ml)$^{-1}$ x day$^{-1}$)] | $\delta$ (day$^{-1}$) | $p$ [(TCID$_{50}$/ml)$^{-1}$ x day$^{-1}$)] | $k$ (day$^{-1}$) | $c$ (day$^{-1}$) | $T_0$ | $V_0$ (TCID$_{50}$/ml) |
|---|---|---|---|---|---|---|---|
| Value | 4.9 x 10$^{-5}$ | 4.2 | 2.8 x 10$^{-2}$ | 3.9 | 4.3 | 4 x 10$^8$ | 4.3 x 10$^{-2}$ |

**Table1.** Estimated parameter values from Baccam et al. [3]

## 4 Results and Discussion

The model as outlined in the previous section, thus parameterized, was used to test the hypothesis.
<u>Test of Hypothesis 2a</u>: Restating, having a high $\tau_1$ would imply an increased virion release count with cost of an increased susceptibility to lysis by Cytotoxic T-

Lymphocytes (CTLs). A lower $\tau_1$ would reduce the susceptibility to CTL mediated lysis at the expense of a reduced virion count.

Result: We observed that the optimal strategy was to increase $\tau_1$ till a threshold (in this particular case it was found to be just less than 4 days). Incidentally, day 4 is also the time at which CTL action is initiated. Hence, the optimal strategy for the virus is to continue the non-productively infected phase till just before CTL initiation. Till CTL action is initiated, the virus will continue to build its population. Increasing $\tau_1$ beyond 4 days would lead to loss of produced virions due to CTL-mediated infected cell lysis. Any decrease below 4 days would reduce the total virus production and hence peak viremia. Hence the optimal control is "bang-bang" (Fig 2.). Bang-bang control strategies have also been known to be optimal in other biological systems like differentiation of B-cells and production of plant seeds [1]. Note that due to the use of a continuous ODE system (which mimics biology more closely) as opposed to a delay-differential equation, some infected cells do burst earlier than day 4.

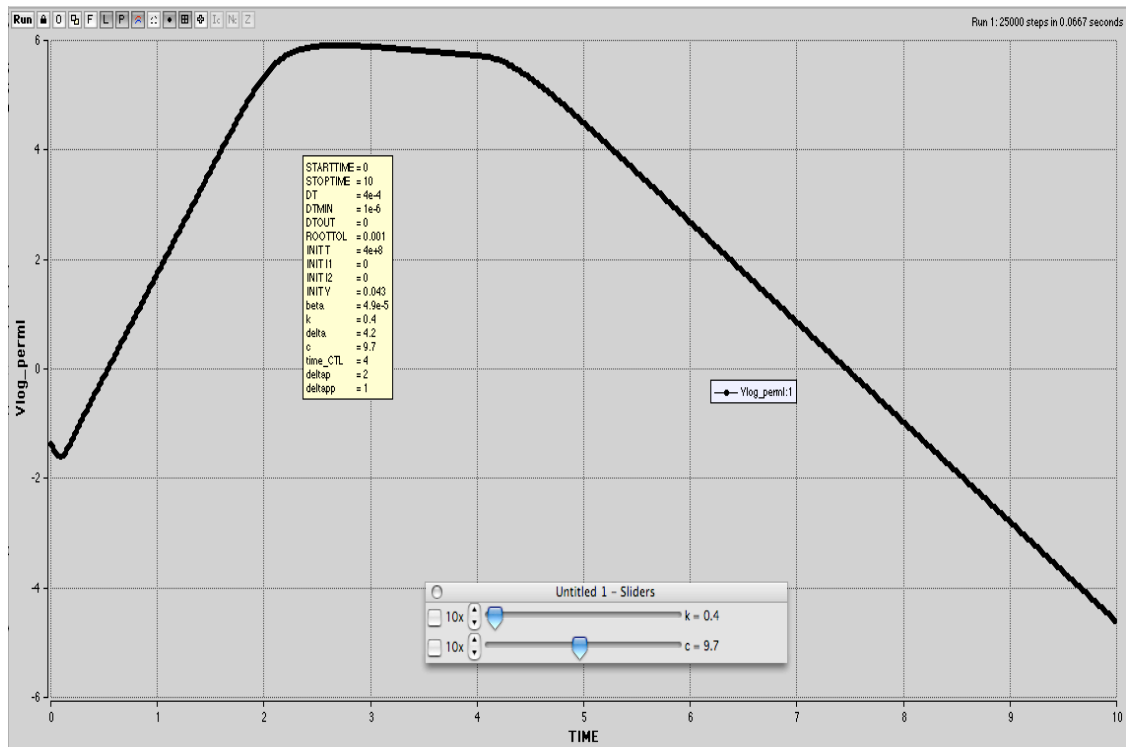

**Fig. 2. Predicted plot of logged viremia ($\log_{10}$ TCID$_{50}$/ml) versus time (days) for Hypothesis 2a**

Test of Hypothesis 2b: Restating, the virus might optimize itself such that it bursts early in the face of a weak antibody response. Conversely, it could burst later (after building up a pool of virions) when confronted with a vigorous antibody response.

Result: The antibody response was varied by manipulation of the virion clearance term $c$ in the ODE system. It was found that the optimal strategy remained conserved under variations in the antibody response i.e. the optimal strategy for the virus was always to burst at $\tau_1 = 4$ days. We can reason about this in the following manner: increasing $\tau_1$ beyond 4 days would lead to loss of produced virions due to CTL-mediated infected cell lysis and any decrease below 4 days would reduce the total virus production and hence peak viremia. Hence antibody response has no effect on $\tau_1$ - a fact that is perhaps not intuitively obvious.

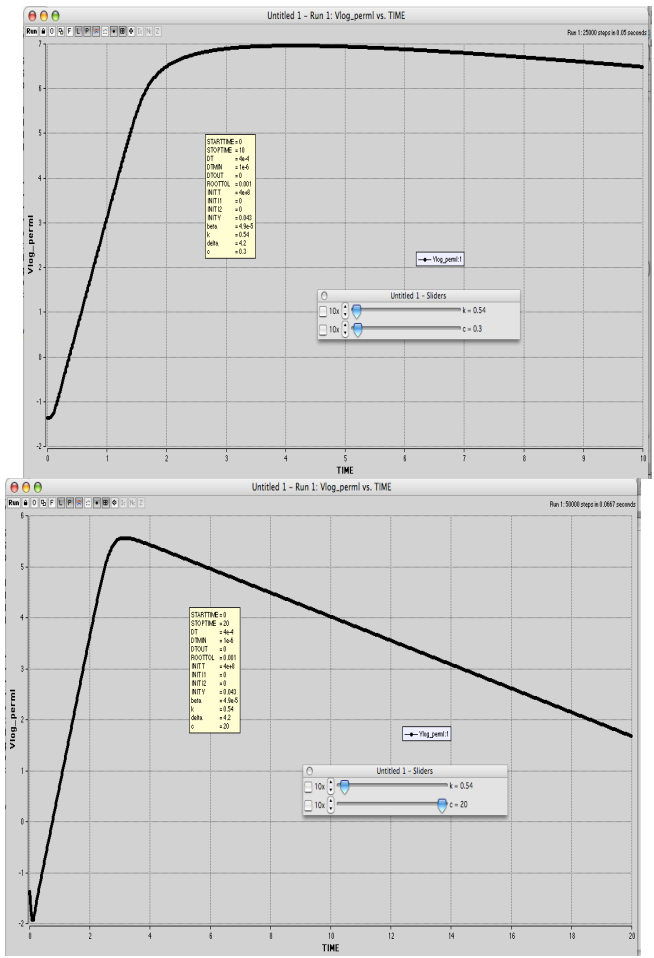

**Fig. 3. Predicted plot of logged viremia ($\log_{10}$ TCID$_{50}$/ml) versus time (days) for Hypothesis 2b (Upper Pane) Low antibody response with $\tau_{1\,=\,}4$ days (Lower Pane) High antibody response with $\tau_{1\,=\,}4$ days. The optimal strategy remains the same (burst just before time to CTL initiation)**

Question 1: Why cannot $\tau_1$ increase indefinitely?

From the preceding discussion, it becomes evident that if $\tau_1$ were to increase indefinitely beyond the time to CTL initiation, then there would be a concomitant decrease in virion output due to CTL-mediated infected cell lysis. Hence the time to CTL initiation sets an upper bound on $\tau_1$.

Question 2: Why is the optimal control "bang-bang"?

Due to physiological limits on cell wall integrity, the time spent in the productively infected phase ($\tau_1$) must be limited. Any attempt to increase it beyond a threshold would merely cause the whole cell wall to break down. Hence, in order to increase virus production, the only "recourse" the virus has is to increase the time spent in the non-productively infected phase and build up the virus population until onset of CTLs. This naturally gives rise to 2 delineated phases ("bang-bang control"). Any intermediate graded response i.e. virion production occurring simultaneously with release is essentially equivalent to the productively infected phase and since the time that can be spent in it is severely limited, we see that it is a sub-optimal strategy. Such strategies are also optimal in diverse biological systems ranging from differentiation of B-cells in the immune system to allocation of energy to seeds in plants [1].

## 4 Conclusions

This work visits virus proliferation from an optimization viewpoint. A few basic assumptions are made: a) the time spent in the productively infected phase is constant and cannot be subjected to optimization beyond a threshold, and b) the virus is trying to optimize virion production and hence peak viremia. Starting from these assumptions, it is posited that the optimal strategy for virus proliferation is to delay burst till onset of Cytotoxic T-Lymphocytes. This so called "bang-bang control" or all-or-none principle is exhibited in many other biological systems like ant colonies and annual plants [1]. However, optimization may not be the only principle at work. In fact, considerations of reliability may be invoked to explain the presence of long-lived latently infected cells (e.g. HIV). These long-lived cells evade detection by CTLs and ensure a prolonged viremia in hosts.

Another conclusion, which is not intuitively obvious, is the fact that the optimal strategy of allocating the maximum time in the non-productively infected phase remains invariant even in the face of a varying antibody response. This strategy is insensitive to the humoral response and depends only on the time to CTL initiation.

The total length of the infected cell lifetime is a measure of "virulence", and a theoretical upper bound has been set on it. Comparing this value to the actual value from field measurements would give us a qualitative understanding of "how far" the virus can still go in optimizing itself e.g. it can be used to determine if the avian-influenza virus is already as virulent as it can be or is it still sub-optimal. In the case of the Influenza A virus from the Baccam et al. study [3], the theoretical upper bound on $\tau_1$ is around 4 days, whereas the observed is around 12 hours, suggesting that the virus is still operating sub-optimally and still has scope to improve by mutating itself. Insights like these could be crucial for bio-surveillance efforts and help inform strategies to cope with future pandemics caused by virus mutations.

Lastly, it is instructive to note that experimental infections of hosts with non-endemic strains (viral strains that have not co-evolved with the host and hence are not operating in an optimal manner) could affect experiment outcome. This would elicit a lower than normal viral response, since the viral strategy would now be characterized by Hypothesis 1 i.e. the time spent in the productively infected phase ($\tau_1$) would just constitute the time required for viral penetration, uncoating of viral core, transcription and assembly and no more.

Clearly more work needs to be done to verify these arguments and an extensive analytical treatment of these arguments coupled with more experimental work will be the subject of future work. The current work highlights the significance of simple mathematical and dynamical models that reveal insights into biological processes as has been done previously in immunology and cell biology [4-13].

**Acknowledgments.** The author wishes to acknowledge fruitful discussions in the 2008 Santa Fe Institute Complex Systems Summer School.

## References

[1] Alan S. Perelson, Majdedin Mirmirani, George F. Oster. 1976. *Optimal Strategies in Immunology: B-Cell Differentiation and Proliferation*. Journal of Mathematical Biology 3, 325-367

[2] Komar, N., S. Langevin, S. Hinten, N. Nemeth, E. Edwards, D. Hettler, B. Davis et al. 2003. *Experimental infection of North American birds with the New York 1999 strain of West Nile virus*. Emerg Infect Dis 9:311-322.

[3] Prasith Baccam, Catherine Beauchemin, Catherine A. Macken, Frederick G. Hayden, Alan S. Perelson. *Kinetics of Influenza A Virus Infection in Humans*. Journal of Virology, Aug. 2006, p. 7590–7599 Vol. 80, No. 15

[4] Soumya Banerjee and Melanie Moses. Scale invariance of immune system response rates and times: perspectives on immune system architecture and implications for artificial immune systems. *Swarm Intelligence*, 2010

[5] Soumya Banerjee, Scaling in the Immune System, PhD Thesis, University of New Mexico, USA, 2013

[6] Soumya Banerjee, Pascal van Hentenryck and Manuel Cebrian. Competitive dynamics between criminals and law enforcement explains the super-linear scaling of crime in cities. *Palgrave Communications,* doi:10.1057/palcomms.2015.22, 2015

[7] Liu, Peng and Calderon, Abram and Konstantinidis, Georgios and Hou, Jian and Voss, Stephanie and Chen, Xi and Li, Fu and Banerjee, Soumya and Hoffmann, Jan Erik and Theiss, Christiane and Dehmelt, Leif and Wu, Yao Wen, A bioorthogonal small-molecule switch system for controlling protein function in cells. *Angewandte Chemie*, 2014

[8] Graessl, M., Koch, J., Calderon, A., S. Banerjee, Mazel, T., Nina Schulze, Jungkurth, J.K., Koseska, A., Leif Dehmelt, Perihan Nalbant. A mechano-sensitive excitable system causes local Rho activity oscillations (in review).
[9] Banerjee S, Levin D, Moses M, Koster F and Forrest S (2011) The value of inflammatory signals in adaptive immune responses. In: Lio, Pietro et al. (eds.) Artificial Immune Systems, 10th International Conference, ICARIS, Lecture Notes in Computer Science, Springer Verlag: Berlin, Germany, vol 6825, pp 1–14.
[10] Soumya Banerjee and Melanie Moses (2009) A hybrid agent based and differential equation model of body size effects on pathogen replication and immune system response. In: P.S. Andrews et al. (eds) Artificial Immune Systems, 8[th] International Conference, ICARIS, 2009, Lecture Notes in Computer Science, Springer Verlag, Berlin: Germany, vol 5666, pp 14–18.
[11] Soumya Banerjee. Analysis of a Planetary Scale Scientific Collaboration Dataset Reveals Novel Patterns. arXiv preprint arXiv:1509.07313, 2015
[12] Soumya Banerjee. An Immune System Inspired Approach to Automated Program Verification. arXiv preprint arXiv:0905.2649. 2009
[13] Soumya Banerjee and Melanie Moses (2010) Modular RADAR: An immune system inspired search and response strategy for distributed systems. In: E. Hart et al. (eds) Artificial Immune Systems, 9th International Conference, ICARIS, 2010, Lecture Notes in Computer Science, Springer Verlag: Berlin, Germany, vol 6209, pp 116–129.
[14] R. I. Macey and G. Oster Berkeley Madonna, version 8.0. Technical report, University of California, Berkeley, California, 2001.
[15] Soumya Banerjee, Model file for Berkeley Madonna for simulations (text file): Optimal Strategies for Virus Propagation, DOI: 10.13140/RG.2.1.3183.4963 (2015)